\documentstyle [12pt,epsfig] {article}

\catcode`@=12
\begin{document}
\thispagestyle{empty}
\begin{flushright} UCRHEP-T317\\September 2001\
\end{flushright}
\vspace{0.5in}
\begin{center}
{\Large \bf Verifiable Origin of Neutrino Mass at TeV Scale\\}
\vspace{1.2in}
{\bf Ernest Ma\\}
\vspace{0.2in}
{\sl Department of Physics, University of California, 
Riverside, CA 92521, USA} 
\vspace{1.5in}
\end{center}
\begin{abstract}\
The physics responsible for neutrino mass may reside at or below 
the TeV energy scale.  The neutrino mass matrix in the $(\nu_e, \nu_\mu, 
\nu_\tau)$ basis may then be deduced from future high-energy accelerator 
experiments.  The newly observed excess in the muon anomalous magnetic moment 
may also be related.
\end{abstract}
\vspace{0.1in}
--------------

\noindent Talk given at the 7th International Workshop on Topics in 
Astroparticle and Underground Physics, Assergi, Italy (September 8-12, 2001)

\newpage
\baselineskip 24pt

\section{Introduction}

The conventional wisdom in neutrino physics is that the origin of neutrino 
mass is at a very high energy scale, say $10^{13}$ GeV or greater, in which 
case there is no hope of verification experimentally.  On the other hand, 
recent papers \cite{marasa,ma01,mara} have shown that it is just as natural 
to have the origin of neutrino mass at the TeV scale and be amenable to 
direct experimental verification.

In the minimal Standard Model with one Higgs doublet $\Phi = (\phi^+,\phi^0)$ 
and 3 lepton doublets $L = (\nu, l)_L$ and singlets $l_R$ only, neutrino 
mass must come from the effective dimension-5 operator \cite{wein,ma98}
\begin{equation}
{1 \over \Lambda} L L \Phi \Phi = {1 \over \Lambda} (\nu \phi^0 - l \phi^+)^2,
\end{equation}
which shows that the form of $m_\nu$ must necessarily be ``seesaw'', i.e. 
$v^2 / \Lambda$ where $v = \langle \phi^0 \rangle$, whatever the underlying 
mechanism for neutrino mass is.

The canonical seesaw mechanism \cite{seesaw} assumes 3 heavy right-handed 
singlet lepton fields $N_R$ with the Yukawa couplings $f L N \Phi$ and the 
Majorana mass $m_N$, hence Eq.(1) is realized with the famous expression
\begin{equation}
m_\nu = {f^2 v^2 \over m_N}.
\end{equation}
Note that lepton number is violated by $m_N$ in the denominator and it should 
be large for a small neutrino mass, i.e.
\begin{equation}
m_\nu \sim \left( {f \over 1.0} \right)^2 \left( {10^{13}~{\rm GeV} \over m_N} 
\right) ~{\rm eV}.
\end{equation}

\section{Higgs Triplet Model}

An equally satisfactory realization of Eq.(1) is to use a Higgs triplet 
\cite{sv,ms} $\xi = (\xi^{++},\xi^+,\xi^0)$ with
\begin{eqnarray}
{\cal L}_{int} &=& f_{ij} [\xi^0 \nu_i \nu_j + \xi^+ (\nu_i l_j + l_i \nu_j)/
\sqrt 2 + \xi^{++} l_i l_j] \nonumber \\ &+& \mu (\bar \xi^0 \phi^0 \phi^0 
- \sqrt 2 \xi^- \phi^+ \phi^0 + \xi^{--} \phi^+ \phi^+) + H.c.
\end{eqnarray}
This model violates lepton number explicitly, but if the parameter $\mu$ is 
set to zero, it becomes the Gelmini-Roncadelli model \cite{gero}, which is 
now experimentally ruled out.  On the other hand, with $\mu \neq 0$ and 
$m_\xi^2$ positive and large, i.e. $m_\xi >> v$,  we have instead \cite{ms}
\begin{equation}
m_\nu = {2 f_{ij} \mu v^2 \over m_\xi^2} = 2 f_{ij} \langle \xi^0 \rangle.
\end{equation}
Note that the effective operator of Eq.(1) is realized here with a simple 
rearrangement of the individual terms, i.e.
\begin{equation}
L_i L_j \Phi \Phi = \nu_i \nu_j \phi^0 \phi^0 - (\nu_i l_j + l_i \nu_j) \phi^+ 
\phi^0 + l_i l_j \phi^+ \phi^+.
\end{equation}
 
Note also that lepton number is violated in the numerator in this case. 
If $f_{ij} \sim 1$, then $\mu/m_\xi^2 < 10^{-13}$ GeV$^{-1}$.  Hence 
$m_\xi \sim 1$ TeV is possible, if $\mu < 100$ eV.  To obtain such a 
small mass parameter, the ``shining'' mechanism of extra large dimensions 
\cite{ahd} may be used.  In that case, the doubly charged 
$\xi^{\pm \pm}$ can be easily produced at colliders and $\xi^{++} \to l_i^+ 
l_j^+$ is a distinct and backgroundless decay which maps out $|f_{ij}|$, 
and thus determine directly the neutrino mass matrix of Eq.~(5) up to an 
overall scale.\cite{marasa}  This model also predicts observable $\mu-e$ 
conversion in nuclei within the sensitivity of proposed future experiments.

\section{Leptonic Higgs Doublet Model}

Another simple and interesting way to have the origin of neutrino mass at 
the TeV scale has recently been proposed.\cite{ma01}  As in the canonical 
seesaw model, we have again 3 $N_R$'s but they are now assigned $L=0$ 
instead of the customary $L=1$.  Hence the Majorana mass terms are 
allowed but the usual $L N \Phi$ terms are forbidden by lepton-number 
conservation.  The $LL\Phi\Phi$ operator of Eq.(1) is not possible and 
$m_\nu = 0$ at this point.

We now add a new scalar doublet $\eta = (\eta^+,\eta^0)$ with $L=-1$, then 
$f L N \eta$ is allowed, and the operator $LL\eta\eta$ will generate a 
nonzero neutrino mass if $\langle \eta^0 \rangle \neq 0$.  The trick now 
is to show how $f \langle \eta^0 \rangle < 1$ MeV can be obtained naturally, 
so that $m_N \sim 1$ TeV becomes possible and amenable to experimental 
verification, in contrast to the very heavy $N_R$'s of the canonical seesaw 
mechanism.

Consider the following Higgs potential:
\begin{eqnarray}
V &=& m_1^2 \Phi^\dagger \Phi + m_2^2 \eta^\dagger \eta + {1 \over 2} 
\lambda_1 (\Phi^\dagger \Phi)^2 + {1 \over 2} \lambda_2 (\eta^\dagger \eta)^2 
\nonumber \\ &+& \lambda_3 (\Phi^\dagger \Phi)(\eta^\dagger \eta) + 
\lambda_4 (\Phi^\dagger \eta)(\eta^\dagger \Phi) + \mu_{12}^2 (\Phi^\dagger 
\eta + \eta^\dagger \Phi),
\end{eqnarray}
where the $\mu_{12}^2$ term breaks lepton number softly and is the only 
possible such term.  Let $\langle \phi^0 \rangle = v$, $\langle \eta^0 \rangle 
= u$, then the equations of constraint for the minimum of $V$ are given by
\begin{eqnarray}
&& v[m_1^2 + \lambda_1 v^2 + (\lambda_3 + \lambda_4) u^2] + \mu_{12}^2 u = 0, 
\\ && u[m_2^2 + \lambda_2 u^2 + (\lambda_3 + \lambda_4) v^2] + \mu_{12}^2 v 
= 0.
\end{eqnarray}
Consider the case $m_1^2 < 0$, $m_2^2 > 0$, and $|\mu_{12}^2| << m_2^2$, then
\begin{equation}
v^2 \simeq -{m_1^2 \over \lambda_1}, ~~~ u \simeq -{\mu_{12}^2 v \over 
m_2^2 + (\lambda_3 + \lambda_4) v^2}.
\end{equation}
Hence $u$ may be very small compared to $v$(= 174 GeV).  For example, if 
$m_2 \sim 1$ TeV, $|\mu_{12}^2| \sim 10$ GeV$^2$, then $u \sim 1$ MeV and
\begin{equation}
m_\nu \sim \left( {f \over 1.0} \right)^2 \left( {1~{\rm TeV} \over m_N} 
\right) ~{\rm eV}.
\end{equation}

Since both $m_N$ and $m_2$ are now of order 1 TeV, they may be produced 
at future colliders and be detected.  (I) If $m_2 > m_N$, then the physical 
charged Higgs boson $h^+$, which is mostly $\eta^+$, will decay into $N$, 
which then decays into a charged lepton and a $W$ boson via $\nu - N$ miixing:
\begin{equation}
h^+ \to l_i^+ N_j, ~~~ N_j \to l_k^\pm W^\mp.
\end{equation}
(II) If $m_N > m_2$, then
\begin{equation}
N_i \to l_j^\pm h^\mp, ~~~ h^+ \to t \bar b,
\end{equation}
the latter coming from $\Phi - \eta$ mixing.  In either case, $m_2$ and $m_N$ 
can be determined kinematically, and $|f_{ij}|$ measured up to an overall 
scale.

In summary, the particle spectrum of the leptonic Higgs doublet model 
consists of the usual Standard-Model particles, including the one physical 
Higgs boson $h_1^0$, 3 heavy $N_R$'s at the TeV scale, and a heavy scalar 
doublet $(h^\pm, h_2^0, A)$ of individual masses $\sim m_2$.  The charged 
Higgs boson $h^\pm$ can be pair-produced at hadron colliders, whereas $N_R$ 
($h^\pm$) can be produced at lepton colliders via the exchange of 
$h^\pm$ ($N_R$).

\section{The Size of Lepton Number Violation}

It has been shown in the above that whereas Majorana neutrino masses have to 
be tiny, the actual magnitude of lepton number violation may come in all 
sizes.

(1) \underline {Large}: $m_N \sim 10^{13}$ GeV in the canonical seesaw 
mechanism.

(2) \underline {Medium}: $|\mu_{12}^2| \sim 10$ GeV$^2$ in the leptonic 
Higgs doublet model with $m_N \sim 1$ TeV.

(3) \underline{Small}: $\mu \sim 10$ eV in the 
Higgs triplet model ($m_\xi \sim 1$ TeV) with a singlet bulk scalar in extra 
large dimensions.

In (2) and (3), direct experimental determination of the relative magnitudes 
of the elements of ${\cal M}_\nu$ is possible at future colliders.

\section{Muon Anomalous Magnetic Moment}

The recent measurement \cite{g-2} of the muon anomalous magnetic moment 
appears to disagree with the Standard-Model prediction \cite{cm} by 
2.6$\sigma$, i.e.
\begin{equation}
\Delta a_\mu = a_\mu^{exp} - a_\mu^{SM} > 215 \times 10^{-11}
\end{equation}
at 90\% confidence level. The origin of this discrepancy may be directly 
related to the TeV physics responsible for neutrino mass.  If the leptonic 
Higgs doublet model \cite{ma01} is combined with a similar model of quark 
masses \cite{ma01_2} to become a supersymmetric model \cite{ma01_3} with 
4 Higgs doublets, then the loop contribution of $\tilde N$ and $\tilde h^+$ 
will in general cause the transition $l_i \to l_j \gamma$.  Hence there 
are predictions \cite{mara} for the muon anomalous magnetic moment as well 
as lepton flavor violating processes such as $\mu \to e \gamma$, $\tau \to 
\mu \gamma$, etc.

If the neutrino mass matrix is hierarchical, then Eq.~(14) implies 
$B(\tau \to \mu \gamma) > 8.0 \times 10^{-6}$, which contradicts the 
experimental upper bound of $1.1 \times 10^{-6}$.  To avoid this restriction, 
the neutrino mass matrix has to be nearly degenerate, in which case we 
have the interesting prediction of 
\begin{eqnarray}
&& {\Gamma (\mu \to e \gamma) \over m_\mu^5} ~:~ {\Gamma (\tau \to e \gamma) 
\over m_\tau^5} ~:~ {\Gamma (\tau \to \mu \gamma) \over m_\tau^5} \nonumber 
\\ &=& {1 \over 2} (\Delta m^2)^2_{sol} ~:~ {1 \over 2} (\Delta m^2)^2_{sol} 
~:~ (\Delta m^2)^2_{atm},
\end{eqnarray}
where bimaximal mixing has been assumed.

In Fig.~(1), the branching fractions of $\tau \to \mu \gamma$ and $\mu \to 
e \gamma$, and the $\mu - e$ conversion ratio in $^{13}Al$ are plotted 
using the lower bound of Eq.~(14), as a function of the common neutrino 
mass $m_\nu$.  The values
\begin{equation}
(\Delta m^2)_{atm} = 3 \times 10^{-3} ~{\rm eV}^2, ~~~ (\Delta m^2)_{sol} = 
3 \times 10^{-5} ~{\rm eV}^2,
\end{equation}
have been chosen according to present data from neutrino-oscillation 
experiments.  At $m_\nu \simeq 0.2$ eV, which is in the range of present 
upper limits on $m_\nu$ from neutrinoless double beta decay, $B(\mu \to e 
\gamma)$ and $R_{\mu e}$ are both at their present experimental upper limits. 
Hence Eq.~(15) will be tested in new experiments planned for the near 
future which will lower these upper limits.

\section{Conclusion}

Physics at the TeV scale may reveal the true origin of neutrino mass, so 
that \underline {accelerator} experiments will become complementary to 
\underline {nonaccelerator} experiments in determining the neutrino mass 
matrix without ambiguity.

\section{Acknowledgements}
This work was 
supported in part by the U.~S.~Department of Energy under Grant 
No.~DE-FG03-94ER40837.

\section{Afterword}
This talk was being given in Assergi, Italy on September 11, 2001, at the 
moment the infernal attack on the World Trade Center in New York began.  All 
of humanity is the victim today and for years to come.

\begin{figure}
\vspace*{13pt}
         \mbox{\epsfig{figure=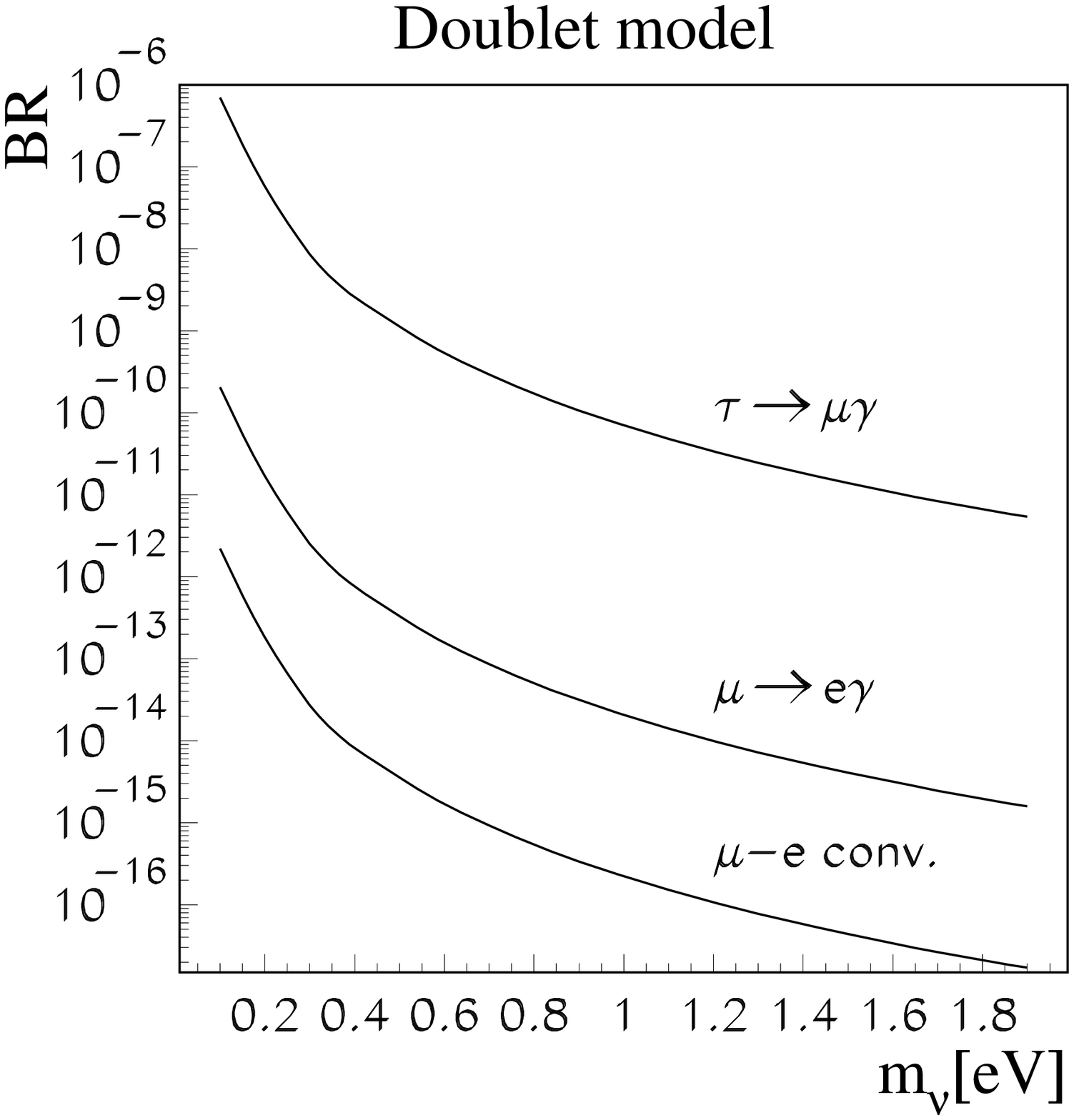,width=15.0cm}}
\caption{Lower bounds on $B(\tau \to \mu \gamma)$, $B(\mu \to e \gamma)$, and 
$R_{\mu e}$ \protect.}
\end{figure}

\end{document}